\documentclass[final,5p]{elsarticle}

\usepackage{graphicx}
\usepackage{hyperref}
\usepackage{color}

\journal{Physics Letter B} 

\begin{document}

\begin{frontmatter}

\title{Probing flavor-dependent EMC effect with W boson production}

\author[a]{W.C.  Chang} 
\author[b]{I.C. Clo\"{e}t}
\author[c]{D. Dutta}
\author[a,d]{J.C. Peng}

\address[a]{Institute of Physics, Academia Sinica, Taipei 11529,
Taiwan}

\address[b]{Department of Physics, University of Washington, Seattle,
Washington 98195, USA}

\address[c]{Department of Physics, Mississippi State University,
Mississippi State, Mississippi 39762, USA}

\address[d]{Department of Physics, University of Illinois at
Urbana-Champaign, Urbana, Illinois 61801, USA}

\date{\today}

\begin{abstract}
A recent theoretical model predicts that the modification of
quark distributions in the nuclear medium (EMC effect) depends on
the flavor of the quarks. We investigate $W$-boson production in 
proton-nucleus collision as a possible tool to test this theoretical
prediction. Several experimental observables in $W$ production
sensitive to the flavor-dependent
EMC effect are identified. Calculations for these experimental
observables at the RHIC and LHC energies are presented using the 
recent flavor-dependent 
EMC model. 
\end{abstract}


\end{frontmatter}

The first definitive evidence for the modification of quark distributions
in the nuclear medium was observed in muon deep inelastic scattering (DIS)
experiment~\cite{emc1}. This surprising finding, called the EMC effect, was
later confirmed by other DIS experiments using electron, muon, 
and neutrinos 
beams~\cite{slac,muons,neutrinos}. Many theoretical models have been proposed
to explain the EMC effect~\cite{geesaman_review,norton_review}. Although these
models are capable of describing certain features of the EMC effect, they
span a wide range of underlying physics. The physics origin of the EMC effect
remains to be better understood.

An effective tool to shed additional light on this subject
is to study the quark flavor dependences of the EMC effect. This was clearly
demonstrated by the measurements of the nuclear dependence of the proton-induced
Drell-Yan process~\cite{drell_yan,dyreview}, which was primarily sensitive to
the $\bar u$ quark distributions in the nucleus. The lack of the nuclear
enhancement of $\bar u$ distributions in these experiments has already
provided strong constraints on various EMC models~\cite{drell_yan}.

To further understand the EMC effect, it would be very
valuable to examine other flavor dependences of the
EMC effect. In this paper, we discuss the possible dependence of the 
EMC effect on the flavor of valence quarks for an $N \ne Z$ nucleus,
where $N$ and $Z$ are the numbers of neutrons and protons in the 
nucleus. We investigate the feasibility for observing such flavor-dependent
EMC effect with $W$-boson production at the RHIC and LHC colliders.

The possibility for a flavor-dependent modification of quark distributions in
$N \ne Z$ nuclei was recently considered by Clo\"{e}t, Bentz, and 
Thomas (CBT)~\cite{ianemc,ian}. In the CBT model, the isoscalar and 
isovector mean fields in a nucleus will modify the quark distributions in the
bound nucleons according to the Nambu-Jona-Lasinio model. These modified 
nucleon quark distributions will then be convoluted with nucleon's
momentum distribution in the nucleus to obtain the nuclear quark distributions.
The strength of the $\rho_0$ field, characterized by the $G_\rho$ coupling,
is determined by the empirical symmetry energy of nuclear matter. The value
of $G_\rho =14.2$ GeV$^{-2}$ is obtained~\cite{ian}.
An interesting consequence of this approach is that the presence of the
isovector vector meson ($\rho^0$) mean field in an $N \ne Z$ nucleus will
modify the $u$ and $d$ quarks in the bound nucleons differently, leading
to flavor dependence of nuclear quark distributions.

\begin{figure}[tbh]
\centering\includegraphics[width=\columnwidth]{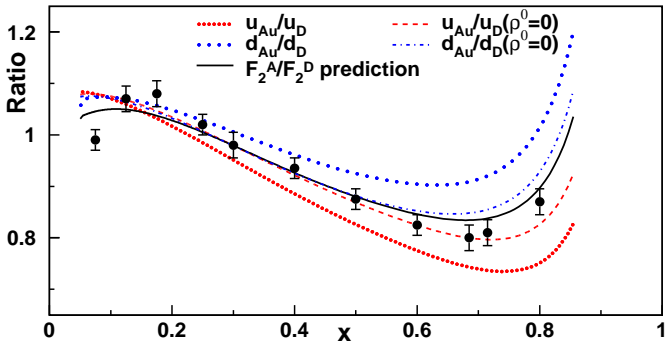}
\caption{Ratios of quark distributions and structure functions in
nuclear matter versus the deuteron plotted as a function of
Bjorken-$x$, at $Q^2 = 10\,$GeV$^2$.
The solid circles are data for $N=Z$ nuclear matter
from Ref.~\cite{day-sick}. The
solid curve is the calculation of $F_2^A/F_2^D$ for $N=Z$ nuclear matter from
Clo\"{e}t, Bentz, and Thomas~\cite{ianemc,ian}. The dotted
curves are the ratios of quark distributions in a gold nucleus to
those in a deuteron, for $u$ and $d$ quarks, respectively. The dashed
and dot-dashed curves are obtained by setting the $\rho^0$ mean field to
0.}
\label{fig1}
\end{figure}

The predictions of the CBT model can be compared with existing DIS data.
Figure~\ref{fig1} shows the $F^A_2/F^D_2$ ratios, where $F^A_2$ 
and $F^D_2$ are the
structure functions (per nucleon) of $N=Z$ nuclear matter and deuteron,
extracted by Sick and Day~\cite{day-sick}. The CBT model calculation,
shown as the solid curve, is in good agreement with the data.
For $N>Z$ nuclei, such as $^{197}$Au and $^{208}$Pb, the
$\rho^0$ mean field causes stronger binding for the $u$ quarks
compared to the $d$ quarks in the protons~\cite{ianemc}. 
Therefore, the nuclear 
modification of the $u$-quark distribution is greater than that of the
$d$-quark distributions. This is illustrated in Fig.~\ref{fig1}, 
where the dotted
and dashed curves show the nuclear modification of $u$ and $d$ quarks,
denoted as $u_{Au}/u_D$ and $d_{Au}/d_D$ 
respectively, in the $^{197}$Au nucleus. Note that for an $N<Z$ nucleus,
the $\rho^0$ mean field will lead to an opposite effect, namely, the 
$d$-quark distribution will be modified more by the nuclear medium than
the $u$-quark distribution.

There are other reasons for different nuclear modifications of $u$ and
$d$ quarks in an $N \ne Z$ nucleus~\cite{kumano}. 
In particular, the well-established
difference between the $u$ and $d$ quark distributions in the proton
would lead to some difference between $u_A/u_D$ and $d_A/d_D$. This `trivial'
flavor dependence is usually taken into account in EMC models automatically
when the nucleon parton distributions are weighted by the $N$ and $Z$
of the nucleus. To illustrate the size of this `trivial' flavor dependence,
the $\rho^0$ mean-field in the CBT model is turned off and the resulting
$u_A/u_D$ and $d_A/d_D$ are shown in Fig.~\ref{fig1}. The relatively small 
difference between $u_A/u_D$ and $d_A/d_D$ confirms that the effect of
this `trivial' flavor dependence is much smaller than that
caused by the $\rho^0$ mean-field.

Since the inclusive DIS on nuclear targets probes the combined nuclear
medium effects of $u$ and $d$ quarks, they do not provide a sensitive
test for the flavor-dependent EMC effect predicted by the CBT model.
Several new measurements sensitive to the flavor-dependent EMC effects
have been considered. They include the parity-violating DIS 
asymmetry proposed at the future JLab 12 GeV facility~\cite{pr12007},
the semi-inclusive DIS on nuclear target first considered by Lu and 
Ma~\cite{ma2006} and recently proposed at JLab~\cite{pr12004}, and
future pion-induced Drell-Yan experiments~\cite{dutta2011}. 
In this paper, we discuss another process, the $W$-boson production 
in proton-nucleus collision, as an experimental tool
sensitive to the flavor-dependent EMC effect. 

\begin{figure}[tbh]
\centering\includegraphics[width=\columnwidth]{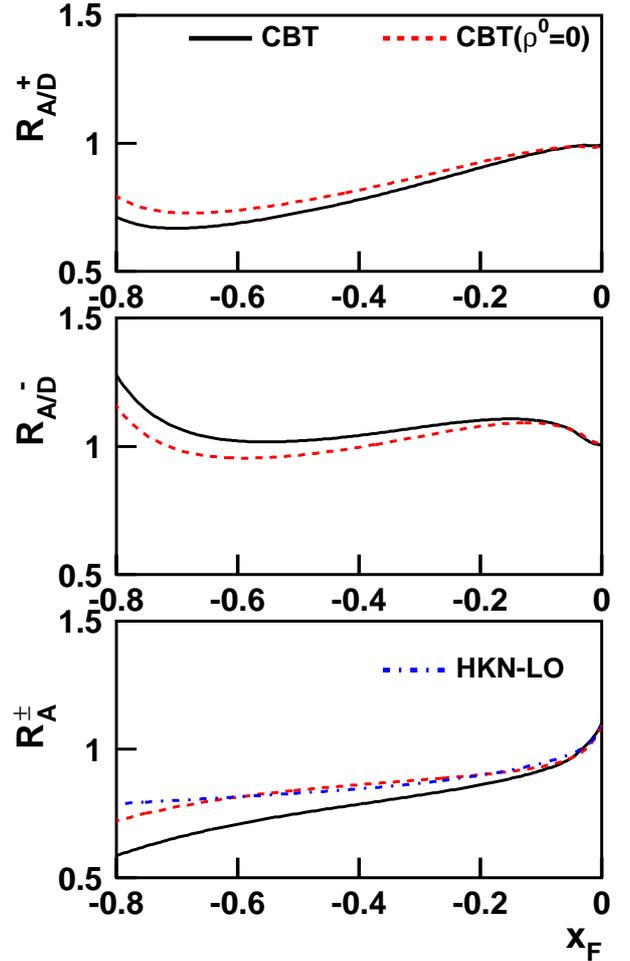}
\caption{Calculations of the three $W$ production ratios, $R^+_{A/D}$,
$R^-_{A/D}$, and $R^{\pm}_A$, for negative $x_F$ at
$\sqrt s = 5.520$ TeV. The solid curves correspond to calculations using
the flavor-dependent PDFs from the CBT model with $N/Z$ equal to that
of $^{208}$Pb. The dashed curves are obtained by setting the $\rho^0$
mean field to zero in the CBT model. The
dot-dashed curve corresponds to calculation using the nuclear PDFs from
Hirai et al.~\cite{kumano07}.}
\label{fig2}
\end{figure}

The differential cross section for $W^+$ production in hadron-hadron collision
can be written as~\cite{barger87}
\begin{eqnarray}
\frac {d \sigma} {dx_F} (W^+) = K \frac {\sqrt 2 \pi} {3} G_F
\left(\frac {x_1 x_2} {x_1 + x_2}\right) \nonumber \\
\left\{ \cos^2 \theta_c[u(x_1) \bar d(x_2) + \bar d(x_1) u(x_2)] + \right. \nonumber \\
\left. \sin^2 \theta_c[u(x_1) \bar s(x_2) + \bar s(x_1) u(x_2)] \right\},
\end{eqnarray}
\noindent where $u(x), d(x),$ and $s(x)$ signify the up, down, 
and strange quark
distribution functions in the hadrons.  $x_1,$ and $x_2$ are the
fractional momenta carried by the partons in the colliding 
proton and nucleus, respectively, and
$x_F = x_1 - x_2$. The $W$ mass, $M_W$, is related to $x_1$,
$x_2$ and the center-of-mass energy squared $s$ as $M_W^2 = x_1 x_2 s$.
$G_F$ is the Fermi coupling constant and $\theta_c$
is the Cabbibo angle. The factor $K$ takes into account the
contributions from first-order QCD corrections
\begin{eqnarray}
K \simeq 1 + \frac {8\pi} {9} \alpha_s(Q^2).
\end{eqnarray}
\noindent At the $W$ mass scale, $\alpha_s \simeq 0.1158$ and $K \simeq
1.323$.  This indicates that higher-order QCD processes are relatively
unimportant for $W$ production.  An analogous expression
for $W^-$ production cross
section is given as
\begin{eqnarray}
\frac{d \sigma} {dx_F} (W^-) = K \frac{\sqrt 2 \pi} {3} G_F
\left(\frac{x_1 x_2} {x_1 + x_2}\right) \nonumber \\
\left\{ \cos^2 \theta_c \,[\bar u(x_1) d(x_2) +  d(x_1) \bar u(x_2)] + \right. \nonumber \\
\left. \sin^2 \theta_c \,[\bar u(x_1) s(x_2) + s(x_1) \bar u(x_2)] \right\},
\end{eqnarray}

To explore the sensitivity of $W$ production to
a flavor-dependent EMC effect, we first consider the ratio
of $W$ production cross sections for $p+A$ and $p+D$ collisions.
If one ignores the much smaller contribution from the strange
quarks, the ratio can be written as 
\begin{eqnarray}
R^+_{A/D} (x_F) & \equiv & \frac {\frac {d\sigma}{dx_F} (p+A \to W^+ + X)}
{\frac {d\sigma}{dx_F} (p+D \to W^+ + X)} \nonumber \\
& \approx  & \frac {u_p(x_1)\bar d_A(x_2)+\bar d_p(x_1)u_A(x_2)}
{u_p(x_1)\bar d_D(x_2)+\bar d_p(x_1)u_D(x_2)},
\end{eqnarray}
\noindent where the subscripts $p$, $D$, and $A$ refer to the parton
distributions in the proton, deuteron, and nucleus, respectively. 
At large negative $x_F$, $x_1<<x_2$, $\bar d(x_2)$ is negligible
compared to $u(x_2)$, and $R^+_{A/D} (x_F)$ becomes
\begin{eqnarray}
R^+_{A/D} (x_F) \approx \frac {u_A(x_2)}{u_D(x_2)}.
\end{eqnarray}
\noindent Similarly, for $W^-$ production at the large negative $x_F$ region,
we have
\begin{eqnarray}
R^-_{A/D} (x_F) & \equiv & \frac {\frac {d\sigma}{dx_F} (p+A \to W^- + X)}
{\frac {d\sigma}{dx_F} (p+D \to W^- + X)} \nonumber \\
& \approx &  \frac {d_A(x_2)}{d_D(x_2)}.
\end{eqnarray}
\noindent Finally, for the ratio of $W^+$ and $W^-$ production in $p+A$
collision at negative $x_F$ region, we obtain
\begin{eqnarray}
R^{\pm}_A (x_F) & \equiv & \frac {\frac {d\sigma}{dx_F} (p+A \to W^+ + X)}
{\frac {d\sigma}{dx_F} (p+A \to W^- + X)} \nonumber \\ 
& \approx & \frac{\bar d_p(x_1)}{\bar u_p(x_1)}\frac {u_A(x_2)}{d_A(x_2)}.
\end{eqnarray}
\noindent Eqs. (5)-(7) show that these three $W$ production ratios
are sensitive to the flavor dependence of the EMC effect.
With the advent of the $W$ production physics program 
at the RHIC~\cite{phenix,star} and LHC~\cite{cms,atlas}
colliders, the measurements of these $W$ production ratios now become
feasible.

\begin{figure}[tbh]
\centering\includegraphics[width=\columnwidth]{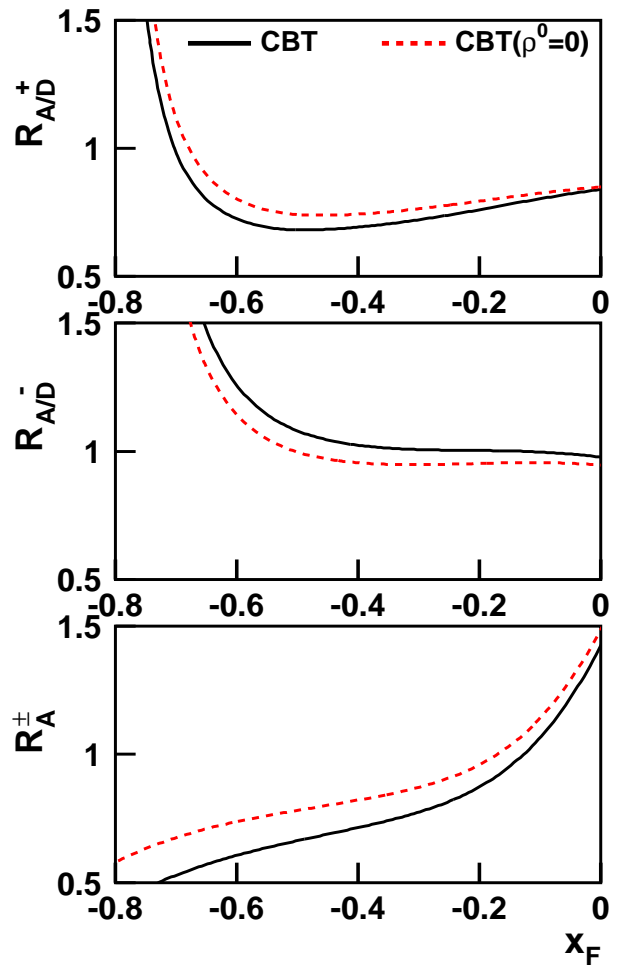}
\caption{Calculations of the three $W$ production ratios, $R^+_{A/D}$,
$R^-_{A/D}$, and $R^{\pm}_A$, for negative $x_F$ at 
$\sqrt s = 200$ GeV. The solid curves correspond to calculations using
the flavor-dependent PDFs from the CBT model with $N/Z$ equal to that
of $^{197}$Au. The dashed curves are obtained by setting the $\rho^0$
mean field to zero in the CBT model.}
\label{fig3}
\end{figure}

\begin{figure}[tbh]
\centering\includegraphics[width=\columnwidth]{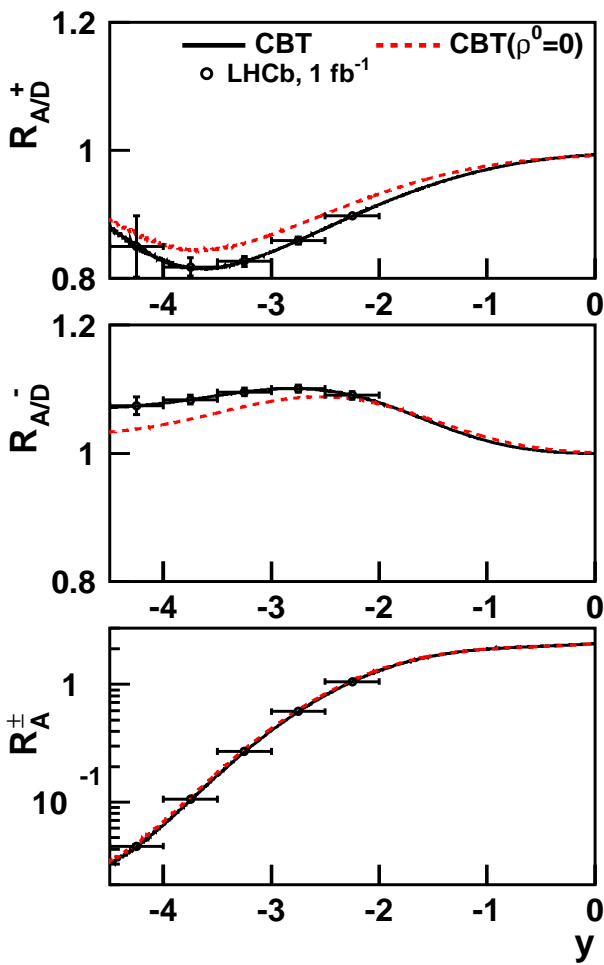}
\caption{Calculations of the three $W$ production ratios, $R^+_{A/D}$,
$R^-_{A/D}$, and $R^{\pm}_A$, for negative rapidity ($y$) of the
charged leptons at
$\sqrt s = 5.520$ TeV. The solid curves correspond to calculations using
the flavor-dependent PDFs from the CBT model with $N/Z$ equal to that
of $^{208}$Pb. The dashed curves are obtained by setting the $\rho^0$
mean field to zero in the CBT model. The expected statistical uncertainties 
for a measurement at the LHCb with an integrated
luminosity of 1 $fb^{-1}$ are also shown.}
\label{fig4}
\end{figure}

To study the sensitivity of the $W$ production in $p+A$ collisions to the
flavor dependence of the EMC effect, we have calculated the three 
ratios ($R^+, R^-,$ and $R^\pm$) 
using the nucleon and nuclear PDFs from the CBT model~\cite{ianemc,ian}.
The PDFs in the CBT model are evolved to the W mass scale using the
DGLAP evolution equations for these
calculations. Instead of using the approximate expressions of 
Eqs. (5)-(7), the full expressions
in Eqs. (1) and (3) are used for the $W$ production cross sections. 
The solid curves in
Fig.~\ref{fig2} show the results for the three $W$ production ratios
for $p+D$ and $p$ + Pb collisions at the LHC energy of $\sqrt s = 5.520$ TeV~\cite{lhcpA}.
Only the results at $x_F < 0$ are shown, since this is the region sensitive
to $u_A(x_2)$ and $d_A(x_2)$ distributions for the valence quarks. 
Also shown in
Fig. 2 are the dashed curves obtained by setting the $\rho^0$ mean field
in the CBT model to zero, that is, removing the flavor-dependent 
EMC effect.
The significant difference in the predicted ratios using the flavor-dependent
versus the flavor-independent nuclear PDFs suggests the feasibility for such
measurements to check the flavor dependence of the EMC effect.
Figure 2 also shows the calculation using the 
flavor-independent nuclear PDFs from Hirai et al.~\cite{kumano07}. 
The good agreement between the calculations using the CBT($\rho^\circ = 0$) 
and the Hirai et al. nuclear PDFs suggests that the systematic uncertainties 
due to the nuclear PDFs are largely cancelled for the ratios $R$ considered 
here.

Figure 3 shows calculations for $W$ production ratios at 
the RHIC energy of $\sqrt s =200$ GeV for $p+D$ and $p$ + Au. The predictions
at the RHIC energy (Fig. 3) are quite similar to those at the LHC
energy (Fig. 2). This is expected from Eqs. 5 - 7, which show that
these ratios are insensitive to the center-of-mass energies. The 
larger $W$ production cross sections at higher energies offer a significant
advantage for measurements at LHC, and we will focus on measurements at
LHC from now on.

Although Figs. 2 and 3 show that the $W$ production ratios are sensitive
to the flavor dependence of the EMC effect, in practice it is not the
$x_F$ distributions of the $W$ which are measured but rather the charged
leptons from the $W$ decays. The relevant ratios can be expressed in
terms of the charged lepton's rapidity 
$y = (1/2) ln[(E+p_l)/(E-p_l)]$, where $E$ and
$p_l$ are the charged lepton's energy and longitudinal momentum, respectively.
We have therefore convoluted the $d\sigma/dx_F$ with the $W \to l \nu$
decay angular distribution, 
$d\sigma / d \cos\theta \sim (1 \pm \cos \theta)^2$, where $\theta$ is the
angle between the lepton $l^\pm$ momentum vector 
and the $W^\pm$ polarization direction
in the $W$ rest frame. From the resulting $d\sigma / d y$, the various
$W$ production ratios can be evaluated as a function of the $y$.

In Fig. 4 we show the predicted $W$ production ratios as a function of
the charged lepton's rapidity at $\sqrt s = 5.520$ TeV using the
nucleon and nuclear PDFs from the CBT model~\cite{ianemc,ian}.  The
dashed curves are obtained by setting the $\rho^0$ mean field to zero
in the CBT model. The differences between the solid and dashed curves
show that the measurement of various $R$ as a function of $y$ remains
sensitive to the flavor dependence of the nuclear modification of
valence quarks. The maximal effect of the $\rho_0$ mean
field is about 3.6\% at y=-3.5 for $R^+_{A/D}$, 3.6\% at y=-4.5
for $R^-_{A/D}$ and 6.5\% at y=-3.8 for $R^\pm_{A}$. 
Although the size of the predicted flavor-dependent EMC 
effect is relatively small, we note that many systematic uncertainties, 
such as those associated with beam luminosity and detector acceptance,
are largely cancelled for the proposed measurement of various ratios.
First results on $W$ production at LHC~\cite{cms, atlas, lhcb} have shown
that high-statistics precision measurements can be obtained with
existing detectors. Although the detector coverage for ATLAS and CMS
does not go beyond $|y|>2.5$, the LHCb has good detector coverage up
to $|y|=4.5$. Figure 4 shows that the expected statistical accuracy
for a measurement at the LHCb with an integrated luminosity of 1
$fb^{-1}$ is sufficient to distinguish the solid from the dashed
curves for $R^+_{A/D}$ and $R^-_{A/D}$. Since the CMS 
and ATLAS experiments have excellent coverage for the mid-rapidity 
region ($-2.5<y<2.5$), the shape of $R$ will be well determined by 
the LHC experiments over a broad rapidity range. This would allow 
a stringent test of any EMC models, including the CBT model.

It is worth noting that this study uses the leading-order (LO) expressions 
for $W$ production. This is necessary since the PDFs in the CBT model was
obtained using the LO formalism. It would be interesting to extend the
calculations to next-to-leading order (NLO) once the corresponding PDFs 
for the CBT model become available. The NLO PDFs 
for the CBT model would
be constrained by the existing extensive DIS data on nuclear targets.
Moreover, future $W$-production cross section data in $pA$ collision
at LHC, including data with isoscalar beams such as $^{12}$C and
$^{40}$Ca, could become available. Data involving isoscalar nuclear 
beams would provide strong additional constraints for the CBT model
when no $\rho^\circ$ mean field is present. Although the absolute
values of the various ratios $R$ could depend on the choice of LO
versus NLO calculations, it is expected that the relative
effect of the $\rho_0$ mean field on $R$ should be insensitive to 
whether LO or NLO calculations are adopted. Therefore, the required
sensitivities of the proposed measurements for identifying the
flavor-dependent EMC effect, as shown in Fig. 4, should also be
applicable in the NLO formualism.

In conclusion, this study shows that $W$ production in $pA$ collisions
is sensitive to flavor dependence of the EMC effect
predicted by the CBT model. We have identified several $W$ production
cross section ratios which are sensitive to the flavor-dependent EMC
effect. The proposed measurements, though quite challenging,
are within the capability of the
existing detectors at LHC and could be carried out in the near
future. A reliable extraction of the flavor-dependent EMC
effect would also require future analysis based on NLO formulation for
the CBT model and nuclear PDFs. The absolute $W$ production cross sections
anticipated for $pA$ collision at LHC would provide important additional
constraints for the NLO analysis. We expect that the proposed measurements, 
together with future theoretical development, could lead to valuable new 
insight on the origin of the EMC effect.

\section*{Acknowledgments}

This work was supported in part by the National Science Council of the Republic
of China and the U.S. Department of Energy and the National Science Foundation.
One of the authors (J.C.P.) thanks the members of the Institute of Physics,
Academia Sinica for their hospitality.

\end{document}